\documentclass[a4paper,onecolumn,11pt]{quantumarticle}
\pdfoutput=1
\usepackage[utf8]{inputenc}
\usepackage[english]{babel}
\usepackage[T1]{fontenc}
\usepackage{amsmath}
\usepackage{hyperref}

\usepackage{tikz}
\usepackage{lipsum}

\begin{document}

\title{Opinion: The simplest quantum computer}
\date{April 1, 2024}

\author{Charles Tahan}
\affiliation{indeterminate}
\orcid{0000-0002-2445-2701}
\email{charlie@tahan.com}
\homepage{http://charlie.tahan.com}
\maketitle

\begin{abstract}
  Instead of writing a review article on the state of the field, I’m going to instead write a retrospective from the year 2040. I’ll tell you how this whole “quantum computing thing” turned out.\footnote{Notice to the literal-minded: the following is a highly speculative and humor-filled guess on the future. Do not assume that any statement in this article is accurate. It solely reflects the personal interests of myself and not reality. Proper referencing is sporadic at best, so please complain about that.}
  \end{abstract}

When I think of impactful opinion pieces on quantum computing, the most obvious are Richard Feynman’s \textit{Simulating physics with computers} \cite{Feynman:1982aa} and David DiVincenzo’s \textit{The Physical Implementation of Quantum Computation} \cite{DiVincenzo:2000aa}. For those of us trying to make working qubits the first decade of this century, the latter governed our lives.

By the year 2024 though, almost everyone’s favorite qubit could meet the core DiVincenzo Criteria. But most important, after two decades of hard-earned progress, at least a dozen groups around the world could make small quantum systems that functioned as fully connected, fully controllable quantum computers. That was a huge accomplishment. We learned that making two qubits undergo a precise entangling operation (a 2-qubit gate) was hard, that integrating lots of qubits together in a well calibrated system was really hard, and that scaling up even thousands of qubits was going to be a mega, well, use your imagination. Everything took way longer than we thought\footnote{For a great laugh, read the ARDA Quantum Computer Roadmap circa 2004 \href{https://qist.lanl.gov/qcomp_map.shtml}{here}.}.

Looking back, I wouldn’t call the progress we made after that \textit{easy}, but a few discoveries propelled us on the path to a useful quantum processing devices. I would summarize them by the following. First, once you learn how to make something work, you learn what you can take away. Second, progress in artificial intelligence served as a forcing function. Both helped us cut corners and choose where to focus our time. What I want to do today is try to summarize how all that discovery played out.

\subsection*{The status of quantum computing in 2024}
First, let’s do the numbers. Variants of the front-running qubit technologies including ion trap qubits, superconducting qubits \cite{PhysRevX.13.031035}, spin-based quantum dot qubits \cite{Stano:2022aa}, and neutral atom qubits \cite{Evered:2023aa} had two-qubit error fidelities of 99-99.9\%, with most single qubit gates better than 99.9-99.99\%. Ion trap quantum computing systems had achieved two-qubit fidelities of 99.9\% in a system of dozens of qubits, with superconducting qubits not far behind (and approaching 100 qubits). Hero devices in some systems had demonstrated fidelities approaching 99.99\%. Primitive logical quantum error correction had been performed in a handful of systems, usually with the aid of post-selection \cite{Bluvstein:2024aa, dasilva2024demonstration}. These demonstrations checked many boxes on the path toward proving that the theories of fault tolerant and quantum error corrected quantum computation were valid. Other qubit systems with long-term promise such as photonic-based or topological qubits were approaching critical demonstrations. Finally, published roadmaps for large-scale quantum computers projected the need for large-scale cryogenic data center class facilities. 

\begin{figure}[t]
  \centering
  \includegraphics[scale=.4]{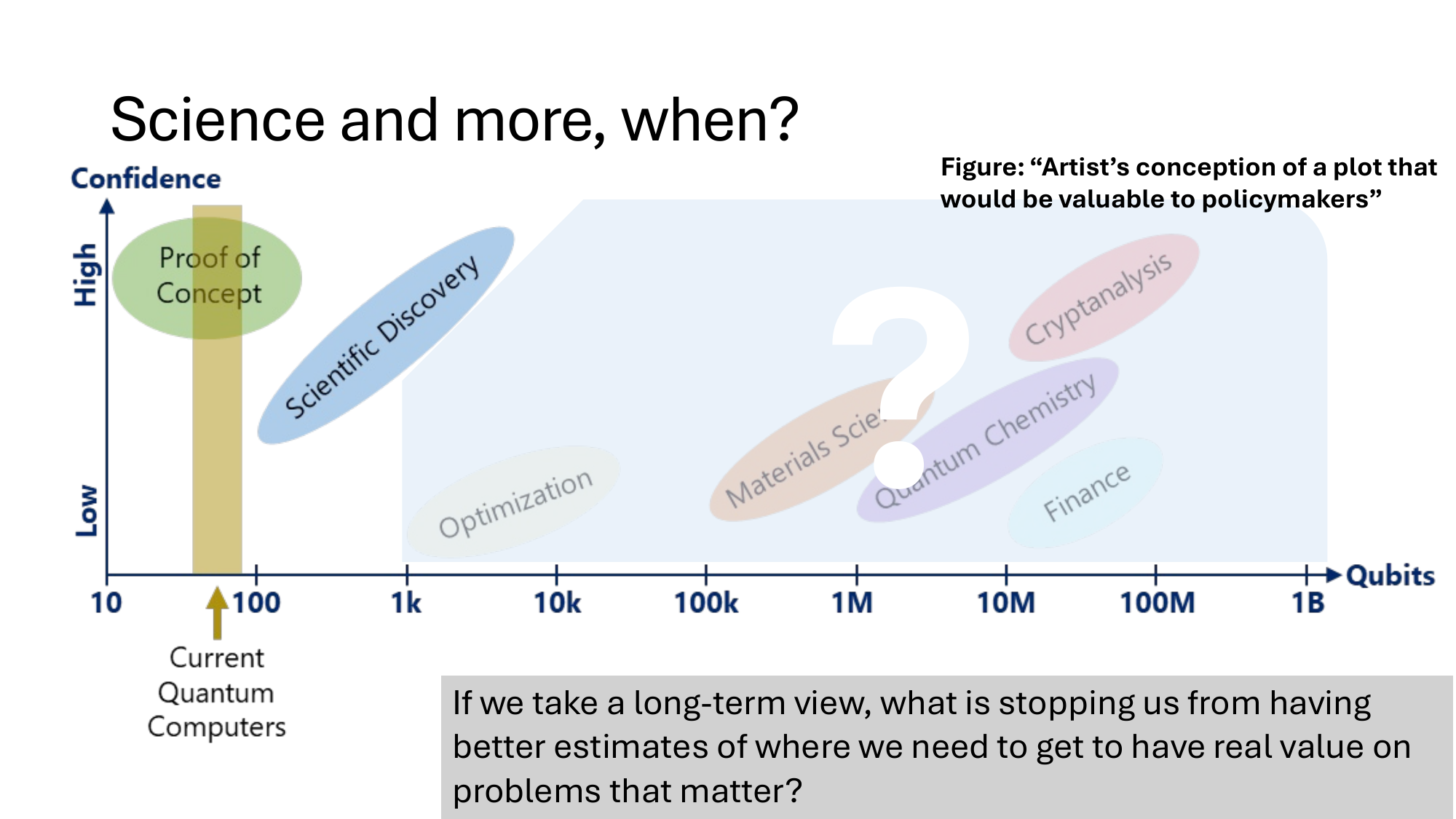}
  \caption{A slide from talks I gave circa 2023.}
  \label{fig:figure1}
\end{figure}

But the most important numbers were the ones we had least confidence in: the resource estimates to run useful algorithms on projected future quantum computers. The most well studied algorithm was Shor's factoring algorithm, which was predicted to break public key cryptography. Resource estimates for qubit count, assuming realistic error-rates and gate speeds of the time, targeted tens of millions or more physical qubits needed \cite{Ekera-RSA2048} to achieve ``well beyond classical'' performance. Concrete---although less optimized---resource estimates for chemistry and materials science \cite{berry2023quantum, Lee:2023aa, Santagati:2024aa, hoefler2023disentangling, beverland2022assessing} pushed quantum computer resource needs beyond even this seemingly insurmountable goal. Other, heuristic-based algorithms gave some people hope for optimization-like problems, as an example, but these algorithms lacked justifiable target requirements (that is, we needed a quantum computer to see if they would work). Other resource estimates for problems in finance \cite{Stamatopoulos2024derivativepricing} or classical machine learning were, frankly, de-motivating, given that it didn't look like they would justify the great expense to build a large quantum computer, although quantum machine learning on quantum data seemed to offer computational separation. It was clear already then that science was the first, if not best, application. I used to show a cartoon in my talks from those days that expressed what we would have loved to know, see Figure~\ref{fig:figure1}.

So, although we had a plan, the path looked very long indeed, with many physics and engineering challenges ahead. So what changed?

\subsection*{All you need is measurement}
\begin{figure}[t]
  \centering
  \includegraphics[scale=.5]{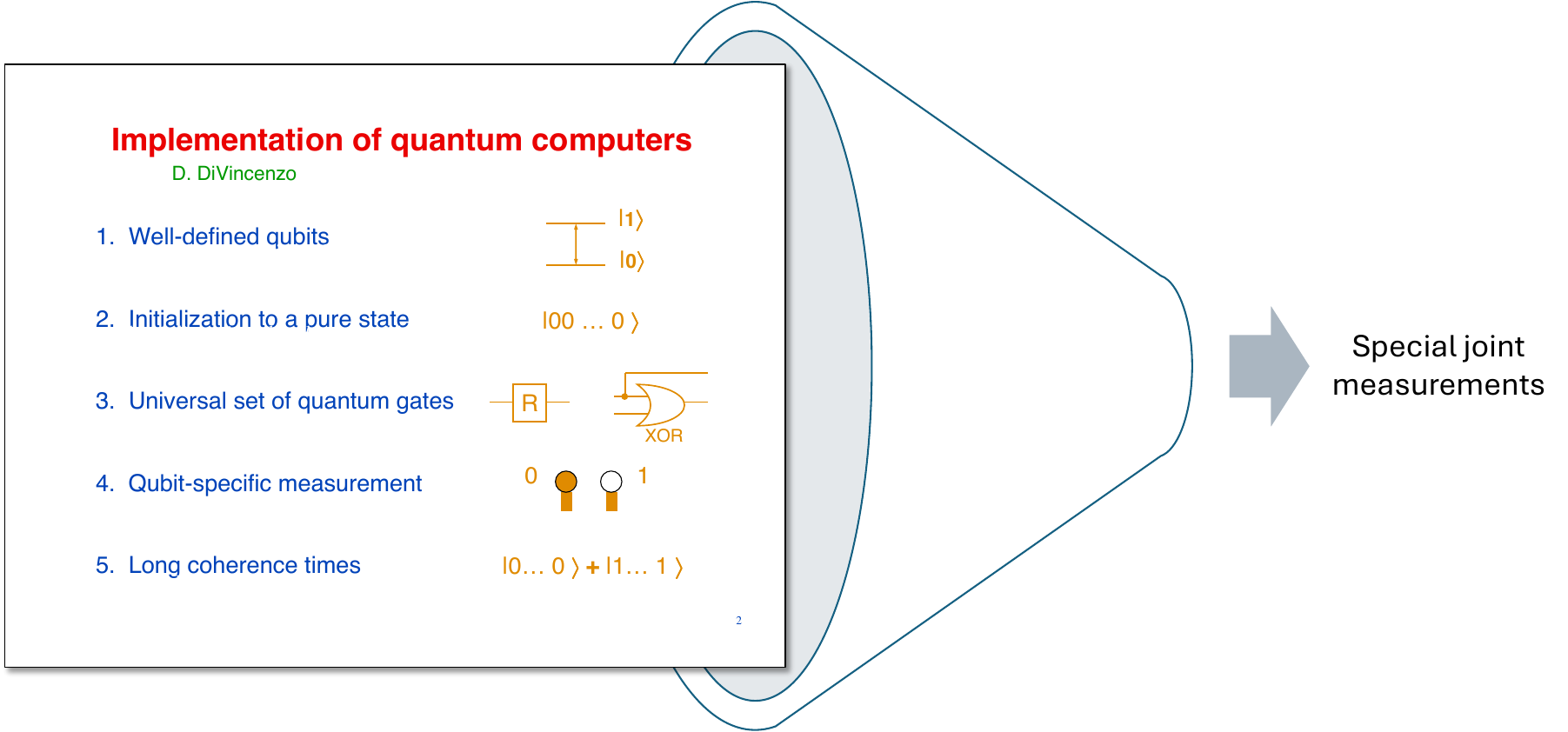}
  \caption{The original DiVincenzo criteria for physical implementations of quantum computers compressed into one un-Vincenzo criteria ("un" for "unified").}
  \label{fig:figure2}
\end{figure}

\textit{Eventually we realized that measurements by themselves were good enough to perform ``good enough'' quantum computation. Although not optimal, this way of thinking unified most solid-state qubits under a common control approach and also unified the necessary conditions for a quantum computer, see Figure ~\ref{fig:figure2}, the so-called ``un-Vincenzo'' criteria. It was actually a big mental shift to get there because generally measurements suck. And it means giving up really good one-qubit gates and still much better (than measurement) two-qubit gates that we spent so much time perfecting!}

Back before qubits worked at all and before we even realized that qubits were hard to make, the community came up with a lot of great ideas \cite{QubitZoo}. These slowly petered back into the collective consciousness as qubits did start to work and work well. One such early idea was the notion of encoding \cite{stean-qec-chapter,PhysRevLett.79.1953}. The scheme involves encoding a qubit of information (an ``encoded qubit'') into a subspace of the larger Hilbert space formed from two to four physical qubits \cite{Lidar2003}. Originally this was proposed to protect the quantum information from global noise sources, but encoding had more practical utility.

Certain types of qubits (like spins, photons, or topological particles) are ``good'' because they don't interact well with the environment. As such, they are often hard to manipulate. For example, for photons, ``dual rail'' encoding defines the qubit itself \cite{PhysRevA.52.3489}. For electron or hole spin qubits in quantum dots (which interact strongly via the Pauli exchange interaction) single qubit gates via a magnetic microwave drive are slow and clumsy. Encoding allows ``all-exchange'' based quantum computing \cite{DiVincenzo:2000ab}, where all gates (except measurements and initialization) are composed of just pair-wise exchange pulses, eliminating the need for microwave generators.

In practice, even for exchange-only qubits, there are physical differences between one and two-qubit gates, even in encoded gates. They have different lengths (number of pulses) for one thing. Coupling  qubits often turns on additional error and leakage channels such as due to charge noise. One-qubit gates are just easier in most systems. Measurement gates generally involve a fundamentally different interaction mechanism (such as spin-to-charge conversion for dots) and are typically much slower and have higher error rates. For reference, a single exchange pulse between two physical dot qubits is 1-10 nanoseconds (among the fastest gates of all qubits) and measurement times were typically 100s of microseconds at the time for 90-99\% error fidelities. That's quite a big difference in relevant parameters.

Nevertheless, when you do the math of quantum error correction (which meant variants of the surface code \cite{PhysRevA.86.032324} then), two-qubit gates set the limits of fault tolerant quantum computation. In other words, optimizing two-qubit gates became the objective, because making them better made everything better. And the community got pretty good at it as I just described. Even though quantum coherent operations require ultra-precise analog pulses operating on extremely delicate analog quantum objects, those were possible in the "NISQ" era \cite{Preskill2018quantumcomputingin}. However, maintaining these stringent error rates while adding qubits involved much more than good electronics. It required very high on/off ratios, reduction of cross talk to make calibration possible, control of drift and yield, and many other systems engineering challenges.

As a general statement, it is usually safe to say that measurement is the worst gate. For some systems, especially atomic-based qubits, measurement could be really, really slow, even milliseconds. Measuring a qubit involves two sequential processes: the quantum portion or “collapse of the wave function” itself, which can be fast. And then converting that quantum signal to classical information, which often requires a long quantum and classical chain of amplifiers, circulators, and electronics, or their equivalent for atomic, molecular, and optical (AMO) systems. However, with respect to the logical quantum error correction cycle, measurement can actually be relatively forgiving. First of all, measurements can be classically amplified by majority voting. Second, erasure type errors were known to have higher quantum error correction breakeven thresholds \cite{Knill:2001aa}. So, for these reasons, mainstream efforts delayed facing the measurement challenges (or improving or exploring alternative measurements) because it was ``good enough'' for such primitive systems and also seemed good enough for future error-corrected systems.

But necessity drove creativity in other qubits, which were forced to make measurements more integral to the quantum gates themselves. The original proposal for Majorana-based topological qubits required dragging anyons around each other \cite{Sarma:2015aa}, an extremely convoluted task, if possible at all in realizable systems. Measurement-based gates offered an alternative \cite{levaillant2015universal}. Similarly, photons don't interact at all, making interacting two photons quite convoluted, unless you "fuse" them together through joint measurements \cite{Bartolucci:2023aa}. Superconducting and super-semi qubits can also be operated this way \cite{Shim:2016aa, PhysRevX.10.041051, yale-dual-rail, PhysRevX.14.011051}, and by analogy other matter qubits. Spins in particular have some nice properties because of their fermionic nature \cite{Freedman2021symmetryprotected, PhysRevB.108.035206, PhysRevApplied.16.064019}. 

Measurements involving multiple qubits at once are needed to drive computation if no entangling gates are used. Because of the monogamy of entanglement, you need special kinds of measurements where some entanglement is left over even after some joint measurement outcomes. Topological qubits are naturally like that, with some overhead. But singlet-triplet measurement (of two quantum dots) are also universal for quantum computation, with even more overhead.

The idea of using measurement to drive quantum computation itself came earlier \cite{Briegel:2009aa}, although in traditional measurement based quantum computing, a large entangled state is presumed at the beginning of the computation. That is not the case for these more gate-focused architectures that are geared to practical realization of fault tolerant quantum computers. 



When you have spent so long getting good at something, it is hard to give it up. And giving up two-qubit gates didn’t make sense in the NISQ era, when we were still learning so much. Nor did it necessarily make sense if you were targeting the full scope of what is possible with a quantum computer. But during this period we learned how to make \textit{systems} simpler, and looking back, the lineage is almost obvious. Encoding let us reduce classical control overhead and make all gates more uniform. Worries about leakage in plain or encoded qubits inspired new gadgets and codes to deal with this error channel, and soon after, more generally, it was discovered that you are better off with biased errors that are the same for all gates \cite{PhysRevLett.120.050505}. Readout for initializing qubits was heavily studied, as initialization could be challenging for some systems. And once we had qubit systems with mid-circuit measurement (\cite{Pino:2021aa} and IBM), we rediscovered the theoretical prediction that measurements plus feed-forward let you create complex entangled states faster than a circuit composed solely of one and two-qubit gates. As true NISQ systems came online, experiments in randomized gates led to deep thinking not only about the spread of entanglement in scrambled systems \cite{Arute:2019aa}, but also about the physics of monitored systems \cite{annurev:/content/journals/10.1146/annurev-conmatphys-031720-030658}. Randomly placed measurements were added in to the mix of random two-qubt gates. It was found that monitored quantum systems have fundamental connection to error correcting codes. It was even discovered that dual measurements can create dynamical codes \cite{Hastings2021dynamically} where the protected logical information moves around the quantum chip, also with implications for heavily measured systems. Again, many of these discoveries came from considering hard to manipulate qubits like topological or photonic quantum computation.

And, begrudgingly, people started to say: "Yeah, I guess you're right, maybe all you need is joint measurements. Let's make them as good as we can, give up on trying to be the best, and not worry about the rest."

So, \textit{the simplest quantum computer} has just one type of gate, and everything is optimized to make that gate high fidelity, repeatable, and fast. But it was worth it, and to understand why I need to explain the second half of the story.

\subsubsection*{Feeding the beast}

\textit{But this realization that joint measurements alone could be universal for quantum computation wasn’t  enough. Because, try as we might, readout was never as good as gates. Even two-qubit gates could be way better, and faster. So it was really hard to give up that functionality, as we weren't really building best or real quantum computers in a sense anymore. The forcing function only came from the success of quantum computing's true competitor: large classical computers in the form of neural networks \cite{doi:10.1126/science.aag2302}, and the subsequent explosion of Large Language Models (LLMs) and the Large Science Models (LSMs) that came after LLMs. In particular, NNs got so good that they seemed to, along with other quantum-inspired algorithms \cite{Tang-phdthesis}, drastically shrink the future value of actual quantum computers.}

Most physical systems aren’t that quantum. They are still quantum, but in a narrow sense governed by the constraints of reality---that most particles have local interactions, that most systems have inherent symmetries \cite{ORUS2014117}. So while nature indeed could solve quantum faster than a classical computer, even it uses tricks. If the the weights of a NN circa 2025 are analogous to a galaxy of stars, one can think of those large number of weights and incorporating many of these tricks inherently, once we learned how to construct them correctly. 

But the Hilbert space created in a true quantum information system is more like a Universe of stars. The most intriguing problems have solutions somewhere within this exponential space. The discovery that quantum-computer-like systems could be used to train classical LSMs to improve their performance was the "one-transistor radio" that gave quantum computing it’s first non-vanity, non-exploratory reason to be economically driven akin to Moore’s law. And the greatest surprise: the models got better at all science, and we didn’t know why. Just like the ingestion of large code-bases caused a step function in the primitive LLMs of the year 2022, some internal structure of quantum mechanics made a qualitative difference to the network's understanding of science itself. It was all the more surprising because the LSMs were not \textit{executing} quantum processing, they were only trained on it.

\begin{figure}[t]
  \centering
  \includegraphics[scale=.5]{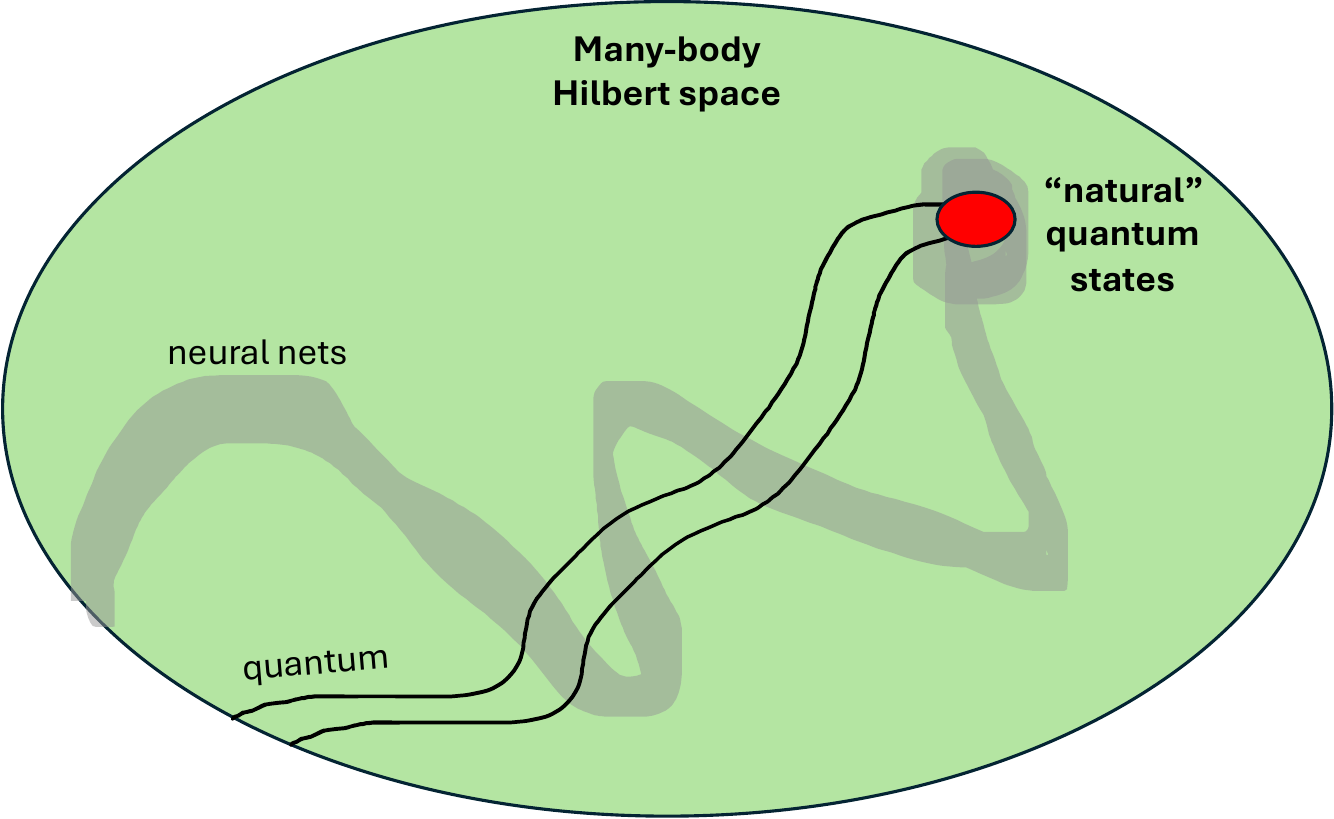}
  \caption{The Hilbert space of a quantum system is exponentially large yet most systems in real life (have constraints (locality, symmetry) that make them occupy a tiny fraction of that space represented by the red region). Quantum processors helped us cut through that space even faster than classical large learning networks.}
  \label{fig:figure3}
\end{figure}

So this sealed the deal. We needed quantum "processors" completely and totally optimized for getting through Hilbert space fast and for getting measurement results out of the system fast. When we hooked them up to what were effectively reinforcement language machines in the biggest high-performance computers, they almost mystically found internal ways to make the quantum coherence (that mattered) last longer. Of course, human hints juiced this process, based on the decade of discoveries before then, such as flag qubits \cite{PhysRevLett.121.050502}, ancilla and Hilbert space expansion, dynamical codes, and more. All these were naturally embedded and emergent in this fabric that could be manipulated purely by measurement. 

This had the convenient hardware implication that there were now both parallel and stepping-stone technologies toward the "best" quantum machine. You want fast, fast measurements that are low power and require the least interconnect and that are preferably compatible with the electronics industry. All these desires favored solid-state qubits. Superconducting, Super-Semi, Spin Quantum Dot, and Topological qubits all fell on this trajectory. Topological can be the best because, even if you ignore their partial resistance to local perturbations, they offer intrinsically lower calibration requirements. But what was really interesting, and critical, is that we found that being perfect in terms of the measurement gates wasn’t strictly necessary, only stability (no drift) was required. The systems that were developed could accommodate variations across the chip and still provide value on the way to better future machines.

And that "solved" the problem, both technical and economic, of quantum computing.

\subsubsection*{Value}

When Elon Musk\footnote{This was before the tragedy involving the neuro-link cyborg hybrid experiment gone wrong.} launched xAI, its goal was to “Discover the answers to life, the universe, everything.” I thought well, duh, the universe runs on quantum mechanics, so a quantum computer would be needed for that. And indeed that turned out to be the case. Quantum computers were the greatest engine for scientific discovery since someone wrote down the modern scientific method.\footnote{No, not our Bacon, the other Bacon.}

Many technologies advance when we give up our preconceived notions of how things ``should'' be done. Quantum was no different. Focusing on the worst thing turned out to be the best thing for the field (and humanity). To the point of how all those details were resolved, what the best n-qubit joint measurements turned out to be, how quantum protection emerged, how large science models were trained, etc. Well, that’s a topic for another retrospective. No need to get into that here! I’ll leave that as an exercise to the reader. 

In the end, I’m glad I spent my career figuring out how quantum could be harnessed for computing instead of say, creating artificial general intelligence. We all know how THAT turned out :-P.

\bibliographystyle{quantum}
\bibliography{simplest.bib}

\onecolumn
\appendix

\end{document}